# VIBESegmentator: Full Body MRI Segmentation for the NAKO and UK Biobank

- Robert Graf* M. Sc. (1,2)
- Paul Platzek m. d. (1)
- Dr. Evamaria Olga Riedel (1)
- Dr. Constanze Ramschütz (1)
- Sophie Starck M. Sc (2)
- Hendrik K. Möller M. Sc (1,2)
- Matan Atad M. Sc (1,2)
- Prof. Dr. Henry Völzke (4)
- PD Robin Bülow (3)
- Prof. Dr. Carsten Oliver Schmidt (4)
- Prof. Dr. Julia Rüdebusch (4)
- Dr. Matthias Jung (5)
- Dr. Marco Reisert (5)
- PD Jakob Weiss (5)
- PD Maximilian T. Löffler (5,1)
- Prof. Dr. Fabian Bamberg (5)
- Prof. Dr. Benedikt Wiestler (1)
- Dr. Johannes C. Paetzold (6)
- Prof. Dr. Daniel Rueckert (7,2)
- Prof. Dr. Jan Stefan Kirschke (1)

*First author / corresponding author

1. Department of Diagnostic and Interventional Neuroradiology, School of Medicine,
   TUM University Hospital
   Neuro-Kopf-Zentrum
   Ismaninger Str. 22
   81675 München, Germany
2. Institut für KI und Informatik in der Medizin, TUM University Hospital
   Technical University of Munich
   Ismaninger Str. 22
   81675 München, Germany
3. Institute for Diagnostic Radiology and Neuroradiology
   University Medicine Greifswald
   Ferdinand-Sauerbruch-Straße
   17475 Greifswald, Germany
4. Institut für Community Medicine, Abteilung SHIP-KEF,
   University Medicine Greifswald,
   Walter Rathenau Str. 48, 5. Etage
   17475 Greifswald, Germany
5. Department of Diagnostic and Interventional Radiology, University Medical Center Freiburg,
   Faculty of Medicine,
   University of Freiburg
   Hugstetter Str. 55
   79106 Freiburg, Germany
6. Department of Radiology, Weill Cornell Medicine,
   1300 York Ave, New York,
   10065, New York, USA
7. Department of Computing
   Imperial College London
   Room 568, Huxley Building
   180 Queen’s Gate
   London SW7 2AZ, UK

# VIBESegmentator: Full Body MRI Segmentation for the NAKO and UK Biobank

**Article Type:** Original Research

**Key Points:**

- **Question:** No completed MRI segmentation model exists that delineates the true transition boundaries of the anatomical structures of bone and muscles.
- **Findings:** We provide a simple-to-use model that automatically segments MRI images, that can be utilized as a backbone for computer aided automatic analysis.
- **Clinical Relevant Statement:** Our segmentation model enables accurate and detailed full-torso segmentation on MRI and CT, improving automated analysis in large-scale epidemiological studies and facilitating more precise body composition and organ assessments for clinical and research applications.

**Abbreviations:**

| Abbreviation | Full Form |
|---|---|
| MR | Magnetic Resonance Tomography |
| CT | Computed Tomography |
| NAKO | German National Cohort |
| UKBB | UK Biobank |
| MedSAM | Medical Segment Anything Model |
| VIBESegmentator | Volume Identification for Bodypart Extraction Segmentator |
| 3D | Three-dimensional |
| FOV | Field of View |
| HASTE | Half Fourier-acquisition Single-shot Turbo Spin Echo |
| VIBE | Volume Interpolated Breath-hold Examination |
| MEVIBE | Multi-echo Volume Interpolated Breath-hold Examination |

# Abstract

**Objectives:** To present a publicly available deep learning-based torso segmentation model that provides comprehensive voxel-wise coverage, including delineations that extend to the boundaries of anatomical compartments.

**Materials and Methods:** We extracted preliminary segmentations from TotalSegmentator, spine, and body composition models for Magnetic Resonance Tomography (MR) images, then improved them iteratively and retrained an nnUNet model. Using a random retrospective subset of German National Cohort (NAKO), UK Biobank, internal MR and Computed Tomography (CT) data (Training: 2897 series from 626 subjects, 290 female; mean age 53±16; 3-fold-cross validation (20% hold-out). Internal testing 36 series from 12 subjects, 6 male; mean age 60±11), we segmented 71 structures in torso MR and 72 in CT images: 20 organs, 10 muscles, 19 vessels, 16 bones, ribs in CT, intervertebral discs, spinal cord, spinal canal and body composition (subcutaneous fat, unclassified muscles and visceral fat). For external validation, we used existing automatic organ segmentations, independent ground truth segmentations on gradient echo images, and the Amos data. We used non-parametric bootstrapping for confidence intervals and Wilcoxon rank-sum test for computing statistical significance.

**Results:** We achieved an average Dice score of 0.90±0.06 on our internal gradient echo test set, which included 71 semantic segmentation labels. Our model ties with the best model on Amos with a Dice of 0,81±0.14, while having a larger field of view and a considerably higher number structures included.

**Conclusion:** Our work presents a publicly available full-torso segmentation model for MRI and CT images that classifies almost all subject voxels to date.

# 1. Introduction

The availability of large-scale biomedical imaging cohorts such as the German National Cohort (NAKO)[1], the Study of Health in Pomerania[2], and the UK Biobank (UKBB)[3] has significantly advanced data-driven research in population health. However, fully leveraging these datasets requires robust, automated methods to extract anatomically meaningful and quantitative information from medical images.[4–8] Whole-body MRI segmentation remains challenging due to complex anatomy, varying contrast across sequences, and the lack of high-quality, standardized reference annotations[9].

Recent deep learning approaches, such as nnUNet[10, 11] and MedSAM[12, 13], have set strong baselines for medical image segmentation. Still,no publicly available method offered comprehensive, voxel-wise segmentation of the full torso in MRI. Although MedSAM [12, 13] is designed as a general-purpose segmentation model, it still faces challenges in achieving broad generalizability. While newer tools like MRISegmenter[14], MRSegmentator[15], and TotalSegmentator-MRI [16] have emerged in parallel to our work, they do not segment every voxel inside the human body and often lack the ability to extend to the border of structures. While avoiding partial volume effects when computing signal values, this causes the organ volume to be underestimated and the 3D shape to be smoothed.

In contrast, we developed the **Volume Identification for Bodypart Extraction Segmentator** (VIBESegmentator), a deep learning-based tool designed to maximize correct tissue voxel coverage, producing segmentation masks that extend to the boundaries of anatomical compartments. This is a feature, critical for accurate volume estimation, spatial mapping, and 3D shape estimation. We utilize a single, unified model across all segmentations, regions and modalities, simplifying deployment and accelerating inference. The development of VIBESegmentator was driven by the demands of large-scale epidemiological studies and the broader need for reproducible, scalable tools in image-based research. Committed to open science, we have made the model and pretrained weights freely available on GitHub (https://github.com/robert-graf/VIBESegmentator). Earlier versions of our model have already been integrated into studies using the UK Biobank[17, 18], supporting cohort-level analysis with high anatomical fidelity.

To further improve generalizability, we have progressively extended support for additional MRI sequences and added compatibility with CT segmentation in the final development phase. To facilitate large-scale medical image segmentation, our model prioritizes full voxel coverage, anatomical precision at compartment boundaries, and multi-modality support.

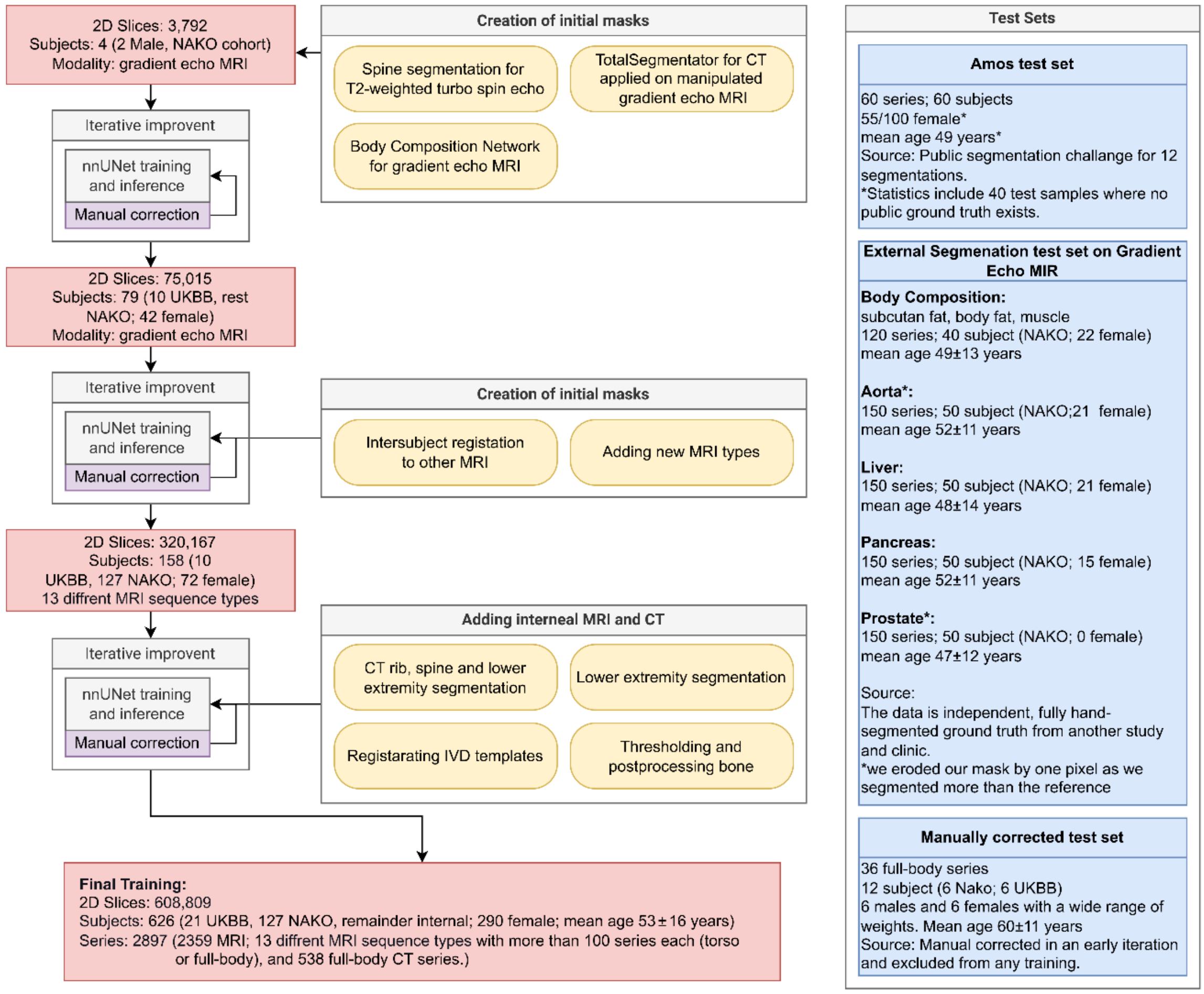

Figure 1: Iterative training procedure: We generated initial masks from different sources (yellow) and then iteratively increased the training dataset (red). Subjects used for training and validation were never used during evaluation: We used four test sets (blue): A manually corrected test set, the Amos dataset, fully independent human annotated segmentation from Freiburg, and a large automatic test set from abdominal organ segmentation. NAKO - German National Cohort; UKBB - UK-Biobank;

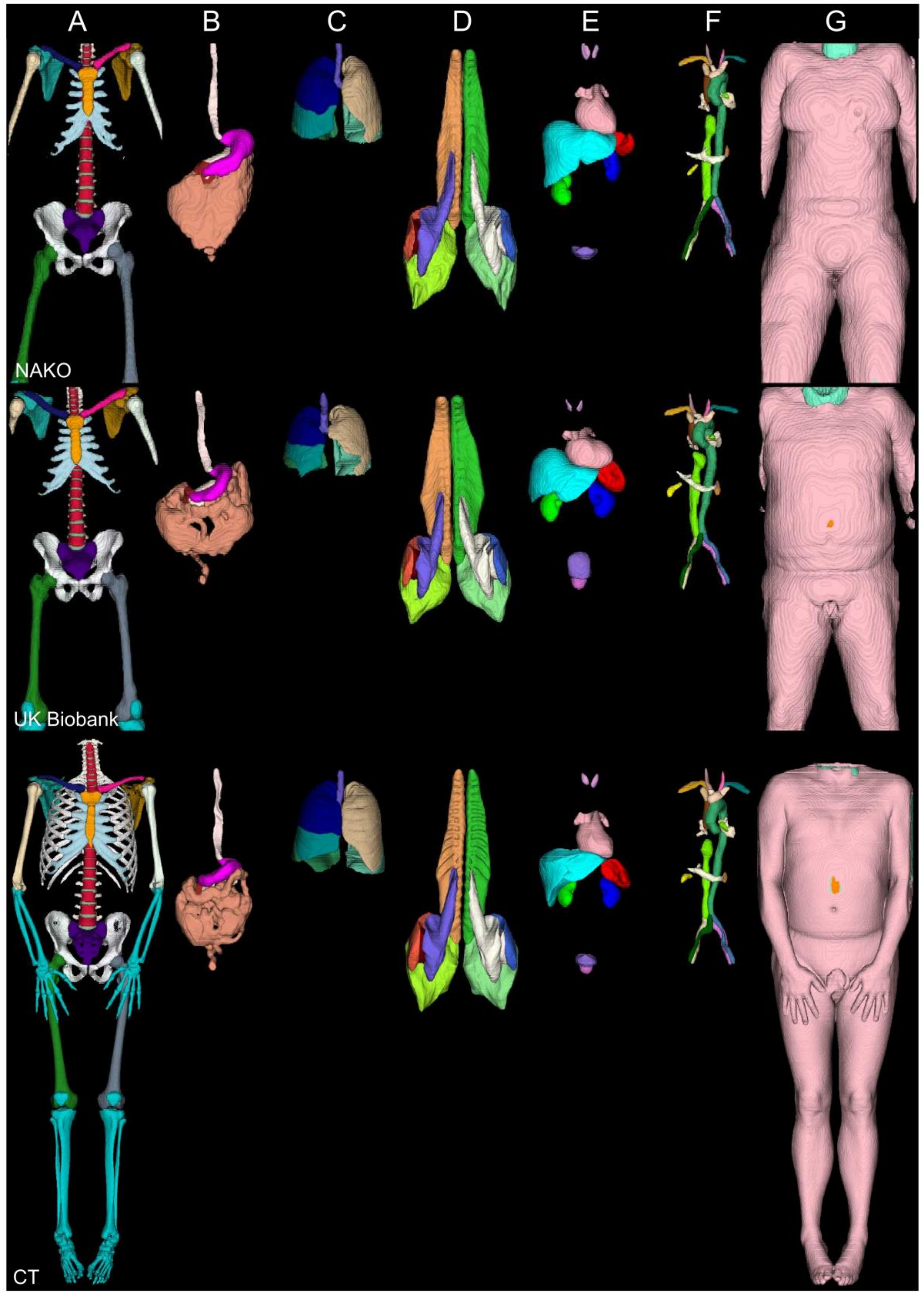

Figure 2: 3D rendering of our segmentations for NAKO, UK-Biobank and internal CT data subjects on Volumetric Interpolated Breath-hold Examination (VIBE) water image. A) bones, interverbal disc, and costal cartilages; B) digestion; C) lung-lobs and trachea; D) muscles; E) organs; F) vessels; G) subcutaneous fat, muscle (other) and visceral fat. NAKO - German National Cohort;

# 2. Materials and Methods

This retrospective study uses a randomized subset of torso imaging data from NAKO, UK Biobank (UKBB) cohorts, as well as internal MRI and CT data for training and evaluation. Ethical approval for data collection was obtained from the respective ethics committees, with informed consent acquired for both NAKO and UKBB. For internal data, we obtained ethical approval from the local ethics committee (593/21 S-NP); informed consent was waived.
We employed an iterative training strategy using the nnUNet framework [10], where newly generated labels were manually corrected and used to improve the model in the next training iteration. We use 3 folds out of the standard 5-fold-cross validation of the nnUNet framework. Our label set is consistent with that of the CT-based TotalSegmentator [19], but we refined the anatomical definitions to better suit MRI data. In particular, we addressed common MRI segmentation challenges, such as under-segmentation at bone and muscle boundaries. Unlike previous approaches, we aimed for touching anatomical borders, deliberately avoiding artificial gaps between regions.

The final dataset comprises stitched gradient echo sequences for water-fat separation [20] [21] from the UKBB and NAKO cohorts, covering a field of view (FOV) from neck to knee. From NAKO, we also included T2-weighted HASTE sequences (torso), proton density images (hip region), six-point Dixon T1-weighted gradient echo scans (abdomen), and sagittal T2-weighted turbo spin echo sequences (spine). To further enhance generalizability, we incorporated an internal dataset consisting of torso MRI, including contrast-enhanced images not present in NAKO, as well as full-body CT scans. All images were resampled to an axial in-plane resolution of 1.4 mm and a through-plane resolution of 3 mm. Our final segmentation model comprised 71 semantic tissue classes for MRI and 72 in CT (Figure 2). We reproduce most of the segmentations that are present in the TotalSegmentator[19]. At the spine, we ignored instance labels and separated all vertebrae into vertebral bodies and posterior elements, as instance segmentations often cause off-by-one errors with nnUNet. We added five additional semantic regions (subcutaneous fat, muscle (not otherwise classified), visceral fat, intervertebral disc, and spinal canal). Due to challenges in annotation, we did not segment ribs in MRI and merged colon with small bowel into a single "intestine" label. Skull and brain segmentations are omitted because they fall outside the FOV and there are already many well-established algorithms available for skull segmentation in MRI[22, 23]. Additionally, structures like the patella, fibula, tibia, ulna, and radius are rarely and only partially captured in the FOV of both NAKO and UKBB datasets, hence categorized into a collective class labeled "bone (other)." We did not reproduce the cysts classes as such pathologies are out of scope for this work and are highly underrepresented in UKBB and NAKO compared to a clinical dataset.

## Ground truth generation

As it is highly time-intensive to fully manually annotate full-torso images to obtain a baseline, we opted for an iterative, stepwise approach, combining available segmentation models with existing ground truth annotations and several iterations of manual refinements (Figure 1). The segmentation refinement process involved four radiologists utilizing ITK-Snap version 3.8.0[24]. Three radiologists (PP, EOR, CR) with 2-3 years of experience corrected the initial segmentation under the supervision of a senior radiologist JSK with 22 years of experience. We used three sources for the initial segmentation proposal: The CT TotalSegmentator on manipulated water images, T2w turbo spine echo for vertebra segmentation from “spineps” model[25–27], and body composition model by Jung et al.[28, 29] to segment the additional subcutaneous fat, muscle (other) and visceral fat. We harmonized the segmentations at their interfaces to ensure seamless segmentation across tissue boundaries. We joined the edges between structures to fill holes in the segmentation, leading to a slight shift from the original ground truths, most noticeable in muscle, visceral fat and bone segmentation. Our body-composition intentionally added the proximal arms, legs, and parts of the shoulder area compared to the reference from Jung et al.[28, 29]. We utilized an iterative approach similar to TotalSegmentator[19], Verse[30], and AbdomenAtlas-8K[31] to increase the number of training images. This iterative process involves correcting predicted segmentation masks and retraining the model until achieving a point of diminishing return: After obtaining the first 4 segmentations, we iteratively trained a nnUNet[10]. For the later stages of model training, we switched to a larger UNet (Residual Encoder UNet-L)[11], as now recommended by nnUNet. After each retraining, we again corrected the new segmentation masks to increase the training dataset. For training, we turned off random flipping to prevent issues predicting left and right labels. Moreover, we employ elastic deformation to generate additional training images to increase the dataset size by a factor of 2 to 500. The segmentation models were trained with random initialization on a single input image where we had three aligned gradient echo images “water”, “in-phase”, or “out-of-phase". We merged the three outputs into a single segmentation. We remove connected components smaller than a specific volume threshold for each organ. If an organ only consists of one tightly packed connected component, we discard all components except the largest. Prioritization during merging focuses on structures prone to under-segmentation, with lower priority segmentations only added to unoccupied voxels. This process aids in eliminating small false positives and underpredictions, particularly for vessels, as reported by TotalSegmentator. In later iterations we added new MRI contrasts. We compared 2 000 of our automated segmentations of inner organs to Kreit et al.[32, 33] on NAKO data to filter for cases where the models disagree (45 subjects with up to 13 different sequences each). This approach identified cases with pathologies in liver, kidney and spleen. Including these, reduced the error in pathological cases, but as numbers were low, and variation is large in pathologies, it did not totally avoid them. In the final iterations we added CT images and improved the bone segmentations with rib[34], spine[30], and lower extremity segmentation[35], as well as by threshold-based inclusion of other bones and template registration to add IVDs. The final training was done on 2359 MRI and 538 CT series for a total of 608,809 resampled 3 mm 2D slices. See Figure 1.

## Evaluations

All evaluations were performed on subjects, not previously included in training and validation datasets. We used four different sources: (1) We compared our model with automated segmentations (n=1000) of liver, kidney, spleen and pancreas from the abdominal model from Kart et al[32, 33], because this was already validated on a large subset of the NAKO. (2) We received independently annotated liver, aorta, pancreas, prostate, visceral fat, subcutaneous fat, and muscle [28, 29] ground truth segmentations from Freiburg on NAKO gradient echo data. This included organ segmentations in 50 subjects and body composition segmentations in 40 subjects, each comprising three spatially aligned images, effectively tripling the size of our internal test set. (3) We validated our model on 36 internally, manually corrected segmentations (3 aligned gradient echo images for a total of 12 subjects), evenly distributed between NAKO and UKBB datasets (50% female), selected to cover a wide range of body weights per sex and dataset. (4) We used the Amos dataset (n=60) [36] to test MRI segmentations of the liver, kidney, spleens, stomach, aorta, gallbladder, inferior vena cava, pancreas, esophagus and adrenal glands. Quality metrics were assessed using the 3D Dice score.

For full body data, we cropped the predictions of MRISegmenter to 50 mm around the ground truth to ensure a fair comparison, as it is not designed for such a large FOV. MedSAM2 received a 2D bounding box derived from the ground truth on the central slice and applied slice propagation.

### Statistical Analysis

For evaluation, we utilized the Scipy [37] and Panoptica Python [38] packages. Nonparametric bootstrapping of 10,000 iterations was utilized to compute 95% confidence intervals (CIs) and Wilcoxon rank-sum test for computing statistical significance.

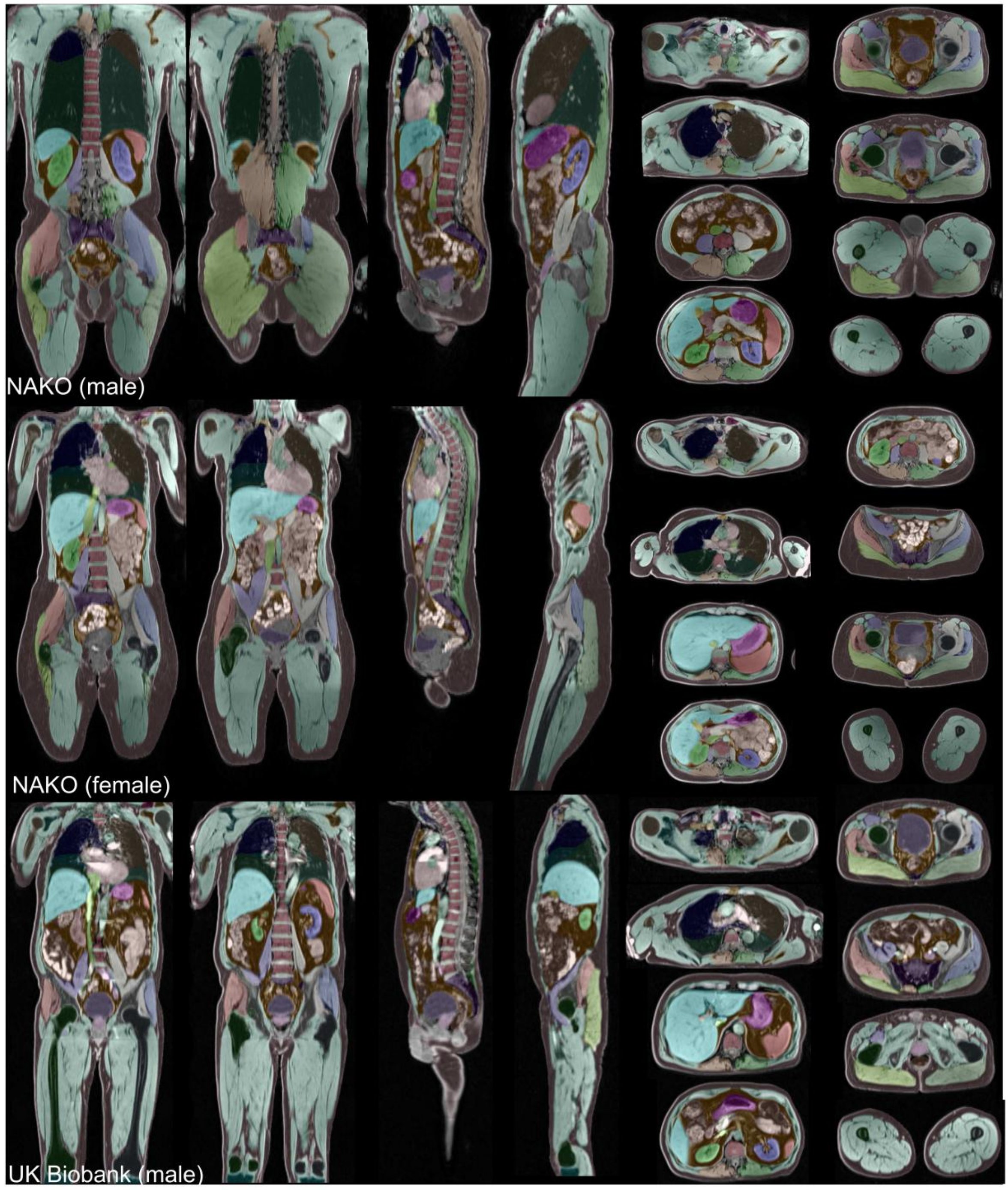

Figure 3: Our segmentation results for three random VIBE images on the NAKO and UK Biobank

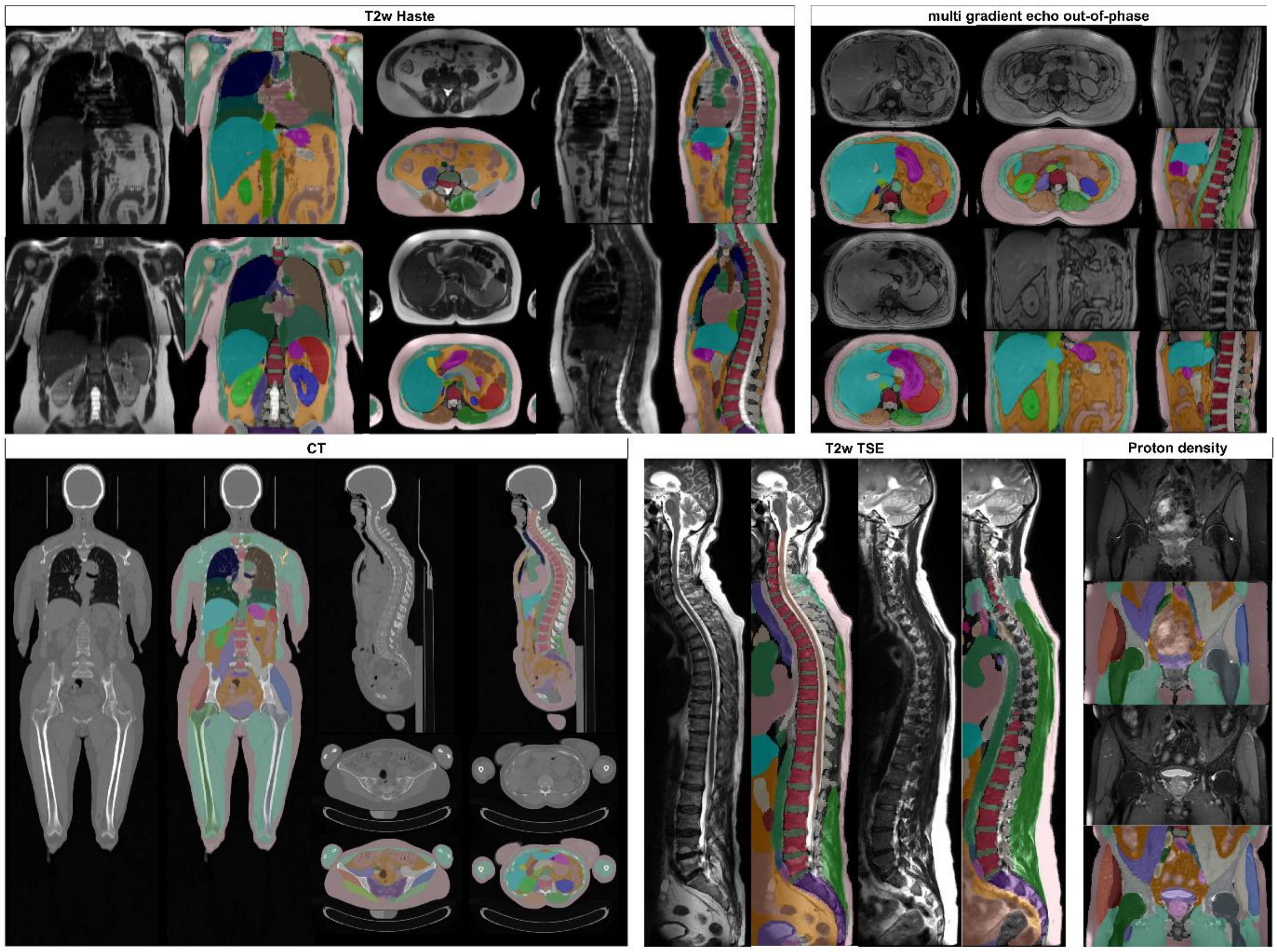

Figure 4: Running our segmentation on other NAKO sequence types and internal CT. We used resampled ground truth segmentation to train on other sequence types. T2w Haste (Half Fourier-acquisition single-shot turbo spin echo) worked best, even with the breathing and heart movement of the images. MEVIBE (multi-echo vibe) has out-of-phase images with clear organ boundaries. The proton density and T2w TSE (turbo spin echo) have signal drops in the NAKO, producing errors where the signal drop is too strong.

# Results

The final training dataset consisted of 608,809 2D slices from 626 subjects, comprising a total of 2,897 series. We provided example segmentations on VIBE sequences (Figure 3) and on other MRI modalities and CT (Figure 4).

## External Evaluations on the model by Kart et al.

In Figure 5, we show Dice scores for all three external evaluations. Our segmentations closely aligned with automatic abdominal segmentations by Kart et al.[32, 33] in case of liver (0.94 Dice; CI: 0.94-0.94), kidney (left/right 0.93/0.94 Dice; CI: 0.93-0.94/0.94-0.94), and spleen (0.93 Dice; CI: 0.92-0.93). The pancreas had a lower Dice score (0.73 ; CI: 0.72-0.73) consistent with Kart et al.[32, 33]. Most discrepancies occur in pathological cases or when the model by Kart et al. undersegments the pancreas in subjects with high visceral fat; an issue not seen in our model. We added a failure analysis through model-agreement with Kart et. Al in the supplemental material.

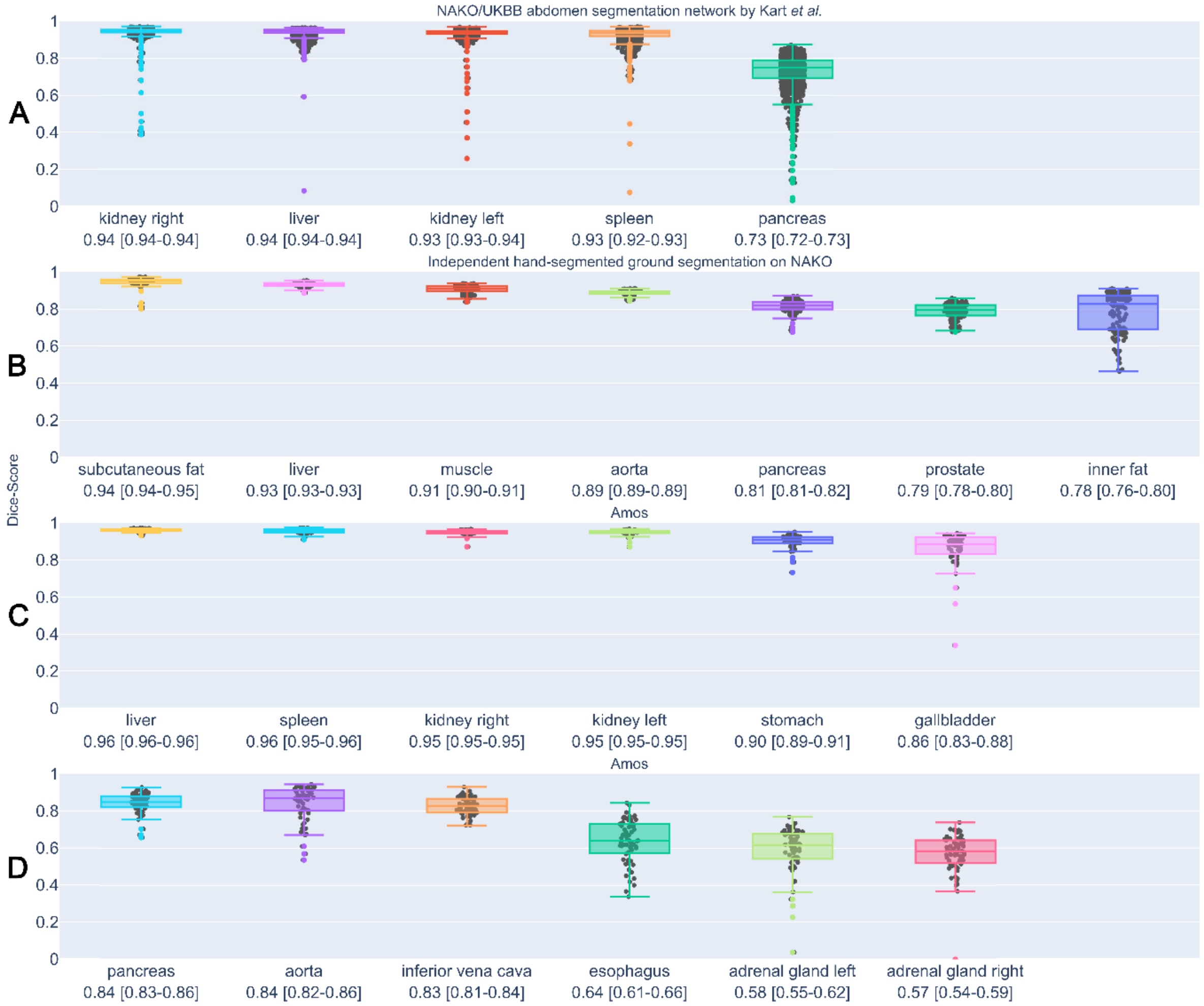

Figure 5: External Validation. A) Dice scores over 1000 test samples compared to the prediction of the abdomen segmentation model by Kart et al.. B) Independent expert segmentations from Freiburg on NAKO images. C) and D) Amos dataset with varying resolution and MR sequences.

## External Evaluations on Gradient Echo Data annotated in Freiburg

We measured the Dice of the subcutaneous fat (0.94 Dice; CI: [0.94, 0.95]) and muscle (0.91 Dice; CI: [0.90, 0.91]) while excluding arm and legs. We extended the visceral fat to match other organ borders, causing our predictions to be changed compared to the reference (0.78 Dice; CI: [0.76-0.80]). Our segmentation of the aorta (0.89 Dice; CI: [0.89, 0.89], erosion by one pixel) included the ascending aorta as well as the vessel wall, unlike the reference that just segmented the lumen of the aortic arch and the descending part. The liver (0.93 Dice; CI: [0.92, 0.93]) and pancreas (0.81 Dice; CI: [0.81, 0.82]) segmentations from Freiburg reached scores that were similar to scores reported by Kart et al on their internal data. In MRI, the capsule of the prostate is usually not segmented, while it is included in CT. We kept the CT delineation, which made our segmentation larger than the independent segmentation from Freiburg (0.79 Dice; CI: [0.79, 0.80], erosion by one pixel). These results were similar to other available segmentation models (Table 2).

## Internal Evaluations on Gradient Echo Data

We compared our model with other multi-structure MRI segmentation models using our internal test dataset (see Table 1). Due to the close alignment in acquisition and segmentation definitions between training and internal test data, higher performance values are expected. On our internal test set, the mean volumetric Dice score across all classes was 0.90 ± 0.06. Individual scores are detailed in Table 1.

We share 44 anatomical structures with TotalSegmentator MRI v. 2.5.0 – D'Antonoli et al. [16], which achieved a mean Dice score of 0.68 ± 0.15 compared to our model's 0.91 ± 0.05 ($p < 0.001$). MRSegmentator – Häntze et al. [15] shares 36 structures, with a mean Dice score of 0.73 ± 0.13 versus our 0.92 ± 0.05 ($p < 0.001$). MRISegmenter – Zhuang et al. [14] overlaps with 32 structures but achieved a substantially lower mean Dice score of 0.39 ± 0.31. This may be due to differences in training data, particularly its smaller field of view and visibly poorer performance on the NAKO/UK Biobank gradient echo data. Several structures—such as the sacrum, gluteus maximus, gluteus medius, iliopsoas, and major vascular structures—were only partially labeled in their reference due to a limited FOV in their training data. Additionally, some classes are not predicted at all in individual gradient echo data. This causes the scores to drop by a large margin. MedSAM2 – Ma et al. [12] showed highly variable performance (mean Dice score 0.42± 0.24), heavily dependent on the selected target organ. Even with manual interaction, MedSAM2 failed to consistently produce accurate segmentations for arbitrary structures in individual images.

Table 1: Dice score and 95% confidence interval (CI) on internal test data of NAKO and UK Biobank Gradient Echo data. Lung lobes combined into “lung right” and “lung left”.

| **Label name** | **Group** | **This study** | D’Antonoli et al. **TotalSegmentator MRI** | Häntze et al. **MRSegmentator** | Zhuang et al. **MRISegmenter** | Ma et al. **Medsam2** |
|---|---|---|---|---|---|---|
| | | *Dice CI:[95%]* | *Dice CI:[95%]* | *Dice CI:[95%]* | *Dice CI:[95%]* | *Dice CI:[95%]* |
| sacrum | bone | **0.95 [0.94-0.95]** | 0.63 [0.61-0.65] | 0.72 [0.71-0.72] | 0.14 [0.11-0.17] | 0.42 [0.39-0.46] |
| humerus left | bone | **0.94 [0.93-0.95]** | 0.71 [0.70-0.73] | N/A | N/A | 0.65 [0.59-0.71] |
| humerus right | bone | **0.94 [0.93-0.94]** | 0.72 [0.71-0.73] | N/A | N/A | 0.64 [0.57-0.70] |
| scapula left | bone | **0.91 [0.91-0.92]** | 0.48 [0.47-0.50] | N/A | N/A | 0.19 [0.18-0.20] |
| scapula right | bone | **0.91 [0.90-0.92]** | 0.48 [0.46-0.49] | N/A | N/A | 0.19 [0.17-0.20] |
| clavicula left | bone | **0.88 [0.86-0.89]** | 0.45 [0.42-0.49] | N/A | N/A | 0.14 [0.11-0.17] |
| clavicula right | bone | **0.88 [0.87-0.89]** | 0.37 [0.33-0.40] | N/A | N/A | 0.19 [0.15-0.23] |
| femur left | bone | **0.96 [0.96-0.97]** | 0.82 [0.81-0.83] | 0.82 [0.82-0.82] | N/A | 0.38 [0.29-0.47] |
| femur right | bone | **0.96 [0.95-0.96]** | 0.83 [0.83-0.84] | 0.83 [0.82-0.83] | N/A | 0.43 [0.34-0.53] |
| hip left | bone | **0.94 [0.94-0.95]** | 0.70 [0.69-0.72] | 0.73 [0.72-0.73] | 0.05 [0.03-0.08] | 0.22 [0.18-0.26] |
| hip right | bone | **0.94 [0.94-0.95]** | 0.67 [0.66-0.69] | 0.73 [0.73-0.74] | 0.05 [0.03-0.07] | 0.21 [0.18-0.25] |
| sternum | bone | **0.86 [0.84-0.88]** | N/A | N/A | N/A | 0.21 [0.18-0.24] |
| stomach | digestion | **0.89 [0.88-0.91]** | 0.82 [0.80-0.84] | 0.84 [0.82-0.86] | 0.68 [0.62-0.74] | 0.40 [0.35-0.46] |
| pancreas | digestion | **0.87 [0.86-0.89]** | 0.62 [0.57-0.67] | 0.68 [0.65-0.70] | 0.45 [0.36-0.54] | 0.37 [0.31-0.43] |
| Esophagus* | digestion | **0.88 [0.87-0.89]** | 0.49 [0.45-0.52] | 0.48 [0.45-0.51] | 0.15 [0.12-0.18] | 0.09 [0.08-0.11] |
| intestine | digestion | **0.92 [0.92-0.93]** | 0.67 [0.64-0.69] | 0.75 [0.73-0.77] | 0.53 [0.50-0.56] | 0.39 [0.34-0.43] |
| duodenum | digestion | **0.85 [0.83-0.87]** | 0.57 [0.52-0.61] | 0.61 [0.56-0.66] | 0.41 [0.34-0.48] | 0.36 [0.30-0.41] |
| adrenal gland right | gland | **0.81 [0.78-0.83]** | 0.56 [0.53-0.59] | 0.57 [0.53-0.60] | 0.39 [0.32-0.46] | 0.24 [0.17-0.31] |
| adrenal gland left | gland | **0.82 [0.80-0.83]** | 0.51 [0.47-0.56] | 0.61 [0.58-0.64] | 0.26 [0.16-0.36] | 0.27 [0.22-0.31] |
| Thyroid gland | gland | **0.76 [0.73-0.80]** | N/A | N/A | N/A | 0.27 [0.18-0.34] |
| gluteus maximus left | muscle | **0.97 [0.97-0.97]** | 0.72 [0.71-0.74] | 0.76 [0.75-0.77] | 0.00 [0.00-0.00] | 0.81 [0.78-0.84] |
| gluteus maximus right | muscle | **0.97 [0.97-0.97]** | 0.72 [0.71-0.73] | 0.78 [0.77-0.79] | 0.00 [0.00-0.00] | 0.82 [0.80-0.83] |
| gluteus medius left | muscle | **0.94 [0.94-0.95]** | 0.73 [0.70-0.75] | 0.78 [0.78-0.79] | 0.08 [0.05-0.11] | 0.70 [0.66-0.73] |
| gluteus medius right | muscle | **0.96 [0.95-0.96]** | 0.76 [0.74-0.78] | 0.82 [0.82-0.83] | 0.02 [0.01-0.03] | 0.71 [0.68-0.74] |
| gluteus minimus left | muscle | **0.93 [0.92-0.94]** | 0.55 [0.53-0.57] | 0.59 [0.57-0.60] | N/A | 0.40 [0.36-0.45] |
| gluteus minimus right | muscle | **0.93 [0.92-0.94]** | 0.57 [0.55-0.59] | 0.62 [0.60-0.64] | N/A | 0.41 [0.37-0.44] |
| autochthon left | muscle | **0.96 [0.96-0.96]** | 0.65 [0.64-0.66] | 0.65 [0.64-0.66] | 0.69 [0.66-0.72] | 0.51 [0.46-0.56] |
| autochthon right | muscle | **0.96 [0.96-0.97]** | 0.64 [0.63-0.66] | 0.65 [0.64-0.66] | 0.70 [0.66-0.73] | 0.56 [0.53-0.59] |
| iliopsoas left | muscle | **0.95 [0.95-0.96]** | 0.77 [0.75-0.78] | 0.72 [0.71-0.73] | 0.43 [0.37-0.48] | 0.63 [0.57-0.68] |
| iliopsoas right | muscle | **0.95 [0.95-0.96]** | 0.75 [0.73-0.76] | 0.71 [0.70-0.72] | 0.34 [0.29-0.39] | 0.57 [0.49-0.63] |
| spleen | organ | **0.93 [0.92-0.94]** | 0.90 [0.89-0.91] | 0.91 [0.90-0.92] | 0.79 [0.72-0.85] | 0.85 [0.82-0.87] |
| kidney right | organ | **0.92 [0.91-0.93]** | 0.89 [0.88-0.90] | **0.92 [0.91-0.93]** | 0.87 [0.84-0.89] | 0.81 [0.77-0.84] |
| kidney left | organ | 0.92 [0.91-0.92] | 0.91 [0.89-0.92] | **0.93 [0.93-0.94]** | 0.87 [0.84-0.89] | 0.83 [0.81-0.86] |
| gallbladder | organ | **0.82 [0.78-0.85]** | 0.56 [0.47-0.65] | 0.57 [0.49-0.65] | 0.58 [0.49-0.66] | 0.54 [0.46-0.60] |
| liver | organ | **0.97 [0.97-0.97]** | 0.92 [0.91-0.92] | 0.94 [0.94-0.95] | 0.89 [0.86-0.92] | 0.85 [0.82-0.87] |
| lung left | organ | **0.97 [0.96-0.97]** | 0.91 [0.90-0.92] | 0.91 [0.90-0.93] | 0.76 [0.73-0.79] | 0.84 [0.81-0.87] |
| lung right | organ | **0.97 [0.97-0.98]** | 0.92 [0.91-0.93] | 0.92 [0.90-0.93] | 0.80 [0.78-0.83] | 0.86 [0.84-0.89] |
| trachea | organ | **0.90 [0.89-0.91]** | N/A | N/A | N/A | 0.10 [0.09-0.13] |
| urinary bladder | organ | **0.94 [0.93-0.95]** | 0.82 [0.78-0.86] | 0.86 [0.84-0.88] | N/A | 0.75 [0.69-0.81] |
| prostate | organ | **0.93 [0.92-0.94]** | 0.74 [0.70-0.78] | N/A | N/A | 0.67 [0.62-0.73] |
| heart | organ | **0.96 [0.95-0.96]** | 0.83 [0.82-0.84] | 0.88 [0.87-0.88] | N/A | 0.81 [0.79-0.82] |
| spinal cord | spine | **0.86 [0.84-0.88]** | N/A | N/A | N/A | 0.12 [0.09-0.17] |
| IVD | spine | **0.87 [0.86-0.89]** | 0.72 [0.70-0.74] | N/A | N/A | 0.24 [0.22-0.26] |
| vertebra body | spine | **0.92 [0.92-0.93]** | 0.81 [0.80-0.82] | 0.70 [0.70-0.71] | N/A | 0.54 [0.49-0.59] |
| Vertebra posterior elements | spine | **0.84 [0.83-0.85]** | N/A | N/A | N/A | 0.15 [0.13-0.17] |
| Spinal channel | spine | **0.89 [0.88-0.91]** | 0.66 [0.64-0.67] | N/A | N/A | 0.13 [0.10-0.16] |
| aorta | vessel | **0.94 [0.93-0.95]** | 0.67 [0.65-0.69] | 0.83 [0.83-0.84] | 0.52 [0.49-0.54] | 0.25 [0.20-0.30] |
| pulmonary vein | vessel | **0.88 [0.87-0.89]** | N/A | N/A | N/A | 0.14 [0.06-0.24] |
| brachiocephalic trunk | vessel | **0.85 [0.83-0.87]** | N/A | N/A | N/A | 0.37 [0.32-0.42] |
| subclavian artery right | vessel | **0.85 [0.83-0.87]** | N/A | N/A | N/A | 0.22 [0.19-0.25] |
| subclavian artery left | vessel | **0.85 [0.83-0.86]** | N/A | N/A | N/A | 0.15 [0.13-0.18] |
| common carotid artery right | vessel | **0.80 [0.78-0.82]** | N/A | N/A | N/A | 0.32 [0.28-0.37] |
| common carotid artery left | vessel | **0.84 [0.82-0.86]** | N/A | N/A | N/A | 0.32 [0.28-0.37] |
| brachiocephalic vein left | vessel | **0.85 [0.82-0.87]** | N/A | N/A | N/A | 0.38 [0.33-0.43] |
| brachiocephalic vein right | vessel | **0.88 [0.85-0.90]** | N/A | N/A | N/A | 0.52 [0.46-0.58] |
| atrial appendage left | vessel | **0.79 [0.76-0.82]** | N/A | N/A | N/A | 0.35 [0.28-0.42] |
| superior vena cava | vessel | **0.91 [0.89-0.93]** | N/A | N/A | N/A | 0.37 [0.34-0.41] |
| inferior vena cava | vessel | **0.91 [0.90-0.92]** | 0.61 [0.56-0.64] | 0.69 [0.67-0.71] | 0.60 [0.55-0.64] | 0.32 [0.28-0.38] |
| portal vein and splenic vein | vessel | **0.80 [0.77-0.83]** | 0.35 [0.32-0.39] | 0.49 [0.45-0.52] | 0.05 [0.03-0.08] | 0.25 [0.20-0.29] |
| iliac artery left | vessel | **0.85 [0.84-0.87]** | 0.54 [0.51-0.58] | 0.60 [0.57-0.62] | 0.09 [0.06-0.12] | 0.20 [0.16-0.24] |
| iliac artery right | vessel | **0.84 [0.83-0.86]** | 0.48 [0.45-0.51] | 0.53 [0.51-0.55] | 0.08 [0.05-0.11] | 0.14 [0.10-0.17] |
| iliac vena left | vessel | **0.89 [0.88-0.90]** | 0.68 [0.65-0.71] | 0.72 [0.71-0.74] | 0.09 [0.06-0.13] | 0.39 [0.33-0.45] |
| iliac vena right | vessel | **0.88 [0.87-0.89]** | 0.66 [0.63-0.69] | 0.72 [0.69-0.74] | 0.10 [0.07-0.14] | 0.30 [0.24-0.35] |
| costal cartilages | other | **0.90 [0.88-0.92]** | N/A | N/A | N/A | N/A |
| subcutaneous fat | other | **0.95 [0.94-0.96]** | N/A | N/A | N/A | N/A |
| muscle rest | other | **0.95 [0.95-0.96]** | N/A | N/A | N/A | N/A |
| inner fat | other | **0.89 [0.88-0.90]** | N/A | N/A | N/A | N/A |

* indicates that we excluded the value for the aggregated results as the reference standard is different. N/A: Not available,

Table 2: Dice score and 95% confidence interval (CI) on Freiburg Gradient Echo images. Each label contains 50 Segmentation with 3 aligned images each.

| **Label name** | **This study** | D'Antonoli et al. **TotalSegmentor MRI** | Häntze et al. **MRSegmentator** | Zhuang et al. **MRISegmenter** | Ma et al. **Medsam2** |
|---|---|---|---|---|---|
| | *Dice CI:[95%]* | *Dice CI:[95%]* | *Dice CI:[95%]* | *Dice CI:[95%]* | *Dice CI:[95%]* |
| liver | 0.93 [0.92-0.93] | 0.93 [0.93-0.94] | **0.94 [0.93-0.94]** | 0.88 [0.87-0.90] | 0.80 [0.75-0.84] |
| pancreas | **0.81 [0.81-0.82]** | 0.77 [0.76-0.78] | 0.74 [0.73-0.75] | 0.43 [0.38-0.48] | 0.73 [0.70-0.76] |
| prostate | 0.79* [0.78-0.80] | **0.80 [0.79-0.81]** | N/A | N/A | 0.48 [0.45-0.50] |
| aorta | **0.89* [0.89-0.89]** | 0.88 [0.88-0.89] | 0.87 [0.86-0.87] | 0.68 [0.68-0.69] | 0.68 [0.66-0.71] |

* Indicates that we used erosion of one pixel, as our reference standard includes the more border.

## External Evaluations on Amos

We also trained on other resampled sequences, as shown in Figure 4. Using the AMOS dataset [36], we demonstrated that our segmentation model can handle multiple MRI modalities, including contrast-enhanced and coronal images, achieving a mean Dice score of 0.81±0.14. See Table 3 for individual results. Among all five currently existing models, MRSegmentator – Häntze et al.[15] (mean Dice: 0.78±0.16), MRISegmenter – Zhuang et al.[14] (mean Dice: 0.81±0.14) and ours (mean Dice: 0.81±0.15) achieved the highest per-class Dice scores. In contrast, TotalSegmentator MRI v. 2.5.0 – D'Antonoli et al.[16] (mean Dice: 0.76±0.15) performed slightly worse than other nnUNet-based models. MedSAM2 – Ma et al. [12] (mean Dice: 0.59±0.28) performed slightly worse than the top three models in segmenting the spleen, pancreas, kidneys, gallbladder, and liver, while its performance on other structures was notably lower. On Amos CT data we achieve an average Dice of 0.74±0.16 compared to 0.81±0.13 Totalsegmentator v. 2.5.0 – Wasserthal. [19]

Table 3: Dice score and 95% confidence interval (CI) on Amos MR images training and test set combined.

| **Label name** | **This study** | D'Antonoli et al. **TotalSegmentator MRI** | Häntze et al. **MRSegmentator** | Zhuang et al. **MRISegmenter** | Ma et al. **Medsam2** |
|---|---|---|---|---|---|
| | *Dice CI:[95%]* | *Dice CI:[95%]* | *Dice CI:[95%]* | *Dice CI:[95%]* | *Dice CI:[95%]* |
| spleen | **0.96 [0.95-0.96]** | 0.91 [0.90-0.91] | 0.93 [0.93-0.94] | **0.96 [0.95-0.97]** | 0.93 [0.92-0.94] |
| kidney right | **0.95 [0.95-0.95]** | 0.92 [0.89-0.94] | **0.95 [0.94-0.95]** | 0.94 [0.93-0.96] | 0.94 [0.93-0.94] |
| kidney left | **0.95 [0.95-0.95]** | 0.91 [0.89-0.92] | 0.94 [0.93-0.95] | **0.95 [0.92-0.96]** | 0.94 [0.93-0.94] |
| gallbladder | **0.86 [0.83-0.88]** | 0.73 [0.66-0.79] | 0.72 [0.66-0.78] | 0.79 [0.73-0.84] | 0.78 [0.73-0.82] |
| esophagus | 0.64 [0.61-0.66] | 0.64 [0.59-0.68] | **0.66 [0.62-0.70]** | 0.61 [0.56-0.66] | 0.26 [0.22-0.29] |
| liver | 0.96 [0.96-0.96] | 0.93 [0.93-0.93] | 0.96 [0.95-0.96] | **0.97 [0.96-0.97]** | 0.93 [0.92-0.94] |
| stomach | **0.90 [0.89-0.91]** | 0.86 [0.83-0.89] | 0.87 [0.84-0.89] | 0.88 [0.84-0.90] | 0.44 [0.37-0.50] |
| aorta | 0.84 [0.82-0.86] | 0.84 [0.82-0.85] | **0.89 [0.87-0.91]** | 0.88 [0.86-0.91] | 0.45 [0.37-0.54] |
| inferior vena cava | 0.83 [0.81-0.84] | 0.74 [0.71-0.76] | 0.83 [0.81-0.84] | **0.86 [0.84-0.88]** | 0.43 [0.38-0.48] |
| pancreas | 0.84 [0.83-0.86] | 0.76 [0.72-0.79] | 0.79 [0.75-0.82] | **0.85 [0.82-0.88]** | 0.62 [0.57-0.66] |
| adrenal gland right | 0.57 [0.54-0.59] | 0.51 [0.47-0.55] | 0.54 [0.50-0.57] | **0.60 [0.56-0.64]** | 0.30 [0.26-0.34] |
| adrenal gland left | 0.58 [0.55-0.62] | 0.56 [0.51-0.60] | 0.53 [0.48-0.58] | **0.62 [0.56-0.67]** | 0.35 [0.31-0.40] |
| duodenum | **0.71 [0.69-0.73]** | 0.56 [0.51-0.60] | 0.58 [0.54-0.62] | 0.68 [0.64-0.71] | 0.29 [0.25-0.33] |

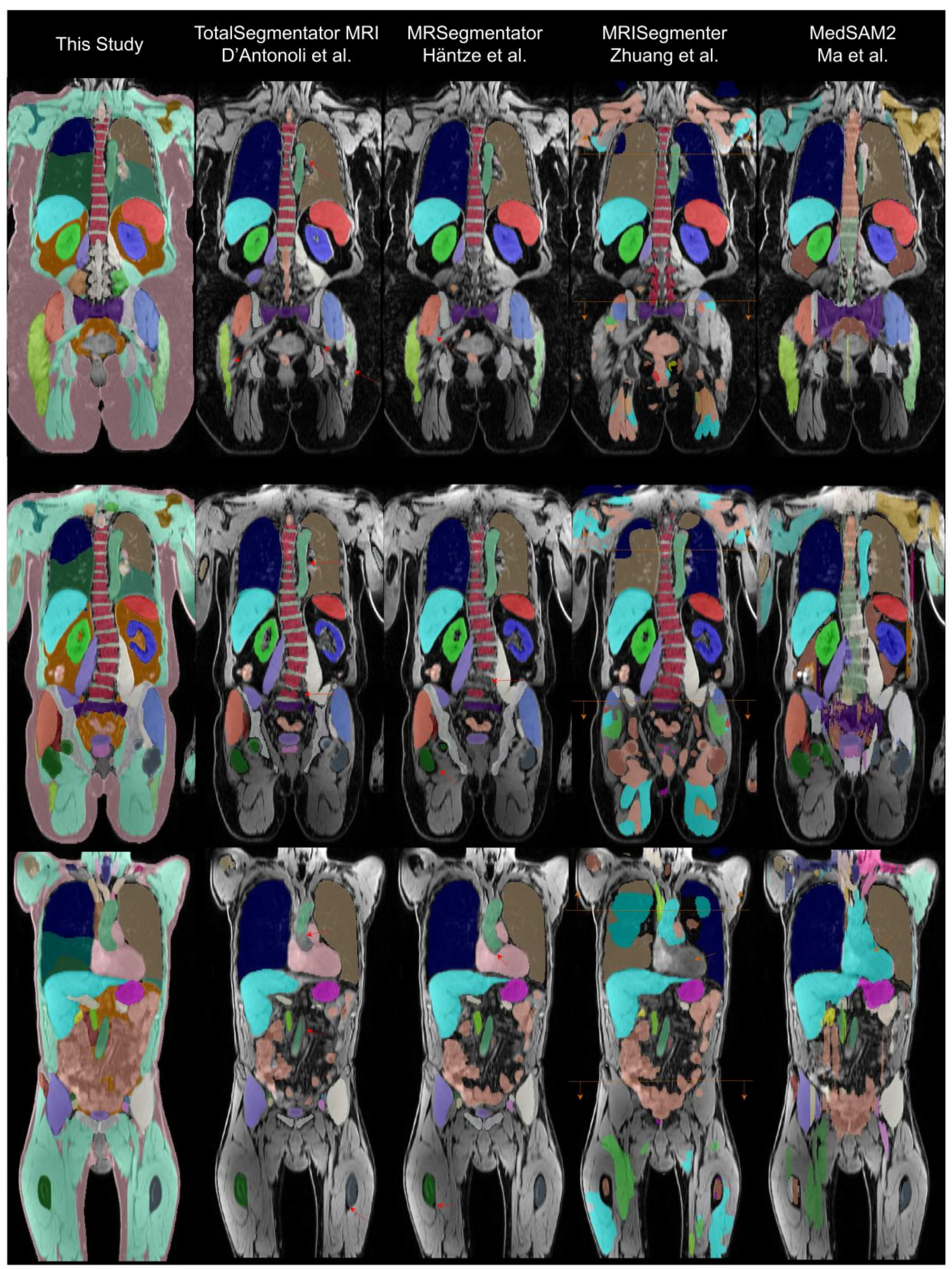

Figure 6: Visual comparison of TotalSegmentator MRI, MRSegmentator, MRISegmenter, MedSAM2, and our method in coronal slices of gradient echo images (water reconstruction) from the NAKO dataset. Our segmentation does not under-segment structures such as muscles and bones and better follows anatomical boundaries. For bone, the dark line is part of the bone structure, as dense bone produces a low proton signal. Examples of under-segmentation are marked with red arrows. We display only those labels that correspond to our label set. Unlike other methods that segment lungs into two sides, our approach identifies all five lung lobes. MRISegmenter was trained on a smaller field of view (FOV); we indicate

its approximate boundary with an orange line. In the final example, MRISegmenter fails to segment the heart and muscles, and MedSAM continued liver segmentation into the heart region (marked with orange arrows). These effects were reproducible in other aligned gradient echo images of the same subject. For visualization only, we merged the individually stored outputs of MedSAM2 predictions. NAKO – German National Cohort.

Table 4: Dice score and 95% confidence interval (CI) on Amos CT images training and test set combined. Our model kept the delineation definition from MR. In CT images there are constraining edges visible for adrenal glands but not from MR images.

| **Label name** | **This study** | Wasserthal et al.<br>**Totalsegmentator CT** | Ma et al.<br>**Medsam2** |
|---|---|---|---|
| | *Dice CI:[95%]* | *Dice CI:[95%]* | *Dice CI:[95%]* |
| spleen | 0.90 [0.89-0.91] | **0.94 [0.93-0.94]** | 0.83 [0.81-0.85] |
| kidney right | 0.92 [0.91-0.92] | **0.94 [0.93-0.95]** | 0.85 [0.84-0.86] |
| kidney left | 0.91 [0.91-0.92] | **0.93 [0.92-0.94]** | 0.85 [0.84-0.86] |
| gallbladder | 0.75 [0.73-0.77] | **0.81 [0.79-0.83]** | 0.56 [0.54-0.58] |
| esophagus | 0.65 [0.63-0.66] | **0.79 [0.78-0.81]** | 0.22 [0.19-0.24] |
| liver | 0.93 [0.92-0.93] | **0.96 [0.96-0.96]** | 0.75 [0.72-0.78] |
| stomach | 0.85 [0.83-0.86] | **0.89 [0.88-0.91]** | 0.40 [0.37-0.42] |
| aorta | 0.85 [0.85-0.85] | **0.92 [0.91-0.92]** | 0.22 [0.19-0.25] |
| inferior vena cava | 0.79 [0.79-0.80] | **0.83 [0.82-0.84]** | 0.20 [0.18-0.23] |
| pancreas | 0.73 [0.72-0.74] | **0.81 [0.79-0.82]** | 0.46 [0.44-0.48] |
| adrenal gland right | 0.49 [0.47-0.50] | **0.70 [0.68-0.71]** | 0.35 [0.33-0.37] |
| adrenal gland left | 0.54 [0.53-0.56] | **0.71 [0.69-0.73]** | 0.34 [0.32-0.36] |
| duodenum | 0.65 [0.64-0.67] | **0.69 [0.67-0.71]** | 0.22 [0.20-0.24] |
| urinary bladder | 0.73 [0.71-0.75] | **0.80 [0.78-0.82]** | 0.56 [0.53-0.59] |
| prostate | 0.45 [0.40-0.49] | **0.46 [0.42-0.51]** | 0.62 [0.59-0.64] |

# 3. Discussion

In this study, we developed a comprehensive full-torso segmentation model incorporating 71 semantic segmentation labels for MR and CT images. Specifically, our model achieves detailed segmentation of 20 organs, 10 muscles, 19 vessels, 16 bones, ribs for CT, and 3 additional spinal structures (intervertebral disc, spinal cord, and spinal canal). Moreover, we include body composition segmentation into three classes, delineating different fat compartments and body parts. [19] Trained on various aligned sequences, our model demonstrates versatility across many different MRI contrasts and CT. Previous studies have highlighted the efficacy of segmentation techniques within the NAKO cohort, such as abdominal organ segmentation and quality control in a large sample of 20,000 participants[32, 33]. Our model extends previous capabilities of whole-body segmentation models to include detailed fat and muscle tissues, building on the work by Jung et al.[28, 29] on VIBE images. This segmentation facilitates fat distribution estimation, akin to the methods employed by Somasundaram and Wu et al. on 6-point Dixon images[39]. Independent of our work, several other full-body MR segmentation frameworks have been developed [14–16]. While these differ fundamentally in their training datasets, our approach also introduces key methodological distinctions. By integrating body composition information, we are able to assign the majority of pixels to specific anatomical or tissue categories, which enhances segmentation stability. A further unique strength of our method lies in its ability to incorporate complex border regions—such as those surrounding bones—and accurately identify contact points between muscle groups and adjacent structures.

After incorporating contrast-enhanced images into our training set, our method achieved performance comparable to MRSegmentator by Häntze et al. [15] and MRISegmenter by Zhuang et al. [14] on the AMOS dataset [36]. In contrast, all models outperformed TotalSegmentor MRI by D’Antonoli et al. [16]. Although MRISegmenter yielded the best results on AMOS, it struggled with gradient echo full-torso images and is limited to a specific field of view. MedSAM2 by Ma et al.[12] performed reliably only for selected anatomical structures, likely due to its limited training scope that excluded certain structures. Moreover, certain anatomical delineations remain inherently ambiguous—such as the precise start and end points of vascular structures—which lack clearly defined boundaries in imaging data, witch a MedSAM like model can only learn by overfitting or manual intervention. These challenges highlight the need for anatomical priors and domain-specific refinement steps in fully automated segmentation systems for MedSAM.

Despite its advantages, the NAKO and UKBB datasets have certain limitations. They predominantly consist of a Caucasian population, making it relatively homogenous in terms of diversity. Whereas the datasets offer a large FOV and high resolution, they contain fewer anomalies, potentially limiting the diversity of pathological cases in the training data. For digestive organs, delineation is more challenging in gradient echo images compared to T1-weighted fast spin echo images with a contrast agent, due to minimal contrast changes in gradient echo images. Although the availability of HASTE, proton density, and MEVIBE sequences is a notable benefit, challenges remain. For instance, very small structures in the UKBB dataset, such as the adrenal gland, are difficult to segment due to being nearly lost in image noise. Additionally, the arms, legs, and head are only partially visible in all MR datasets, leading to undefined segmentation behavior in these regions and necessitating a separate improvement dataset.

In conclusion, our work significantly advanced the field of MRI segmentation by providing a detailed and refined approach to full torso segmentation, particularly in large datasets like UKBB or NAKO. Our model can be used for automatic processing and extraction of structures in MRI and is the most detailed full torso segmentation model to date, allowing for a classification of almost all voxels within the torso.

# 5. Supplementary Material

### Intermediate Failure Analysis

We used the third last iteration and the abdominal segmentation network by Kart et al.[1, 2] to find failure cases to include them in the training. We segmented 1,000 images (448 female, mean age 50) using the abdominal segmentation network[1, 2] and compared our second last model's output. We inspected 40 outliers where at least one segmentation was two standard deviations below the mean Dice score. Four images had a water/fat inversion error in the MR signal. Our segmentation was robust to these errors, producing only minor inaccuracies at the inversion lines. In addition to these issues, both models produced errors for horseshoe kidneys, liver and kidney cysts, and abnormally large kidneys and spleens, which were only partially segmented. Our model produced faults in 12 out of 40 cases, while the abdominal segmentation method[1, 2] produced faults in 35 out of 40 cases. We counted a segmentation as faulty if at least 10% of the organ was missing or over-segmented. We corrected these and added them to the next training iteration. We repeated the experiment with another 1,000 images (421 female, mean age 50). Results showed that the model was more robust to cyst and size changes, and only extreme cases caused segmentation errors after including other images with pathologies. We did not find issues with our pancreas segmentation in both tests, but observed that the abdominal segmentation network[1, 2] has issues with pancreas segmentation if the person has larger amounts of body fat. Known failure cases include subjects with additional or displaced organs, such as with kidneys, or unseen pathologies. Similar issues occur with the head, feet, hands, and forearms, as these are outside the field of view of our training data. We have not observed segmentation errors in bone and lung segmentation in our healthy subjects so far.